%% file: main_camera_ready.tex
\documentclass{article}
\usepackage{spconf,amsmath,graphicx}
\usepackage{arydshln}
\usepackage{multirow}
\usepackage{booktabs}
\usepackage{color}
\usepackage{amsfonts}
\usepackage{pifont}
\usepackage{hyperref}
\usepackage{booktabs}
\usepackage{cite}
\usepackage{scrextend}

\newcommand{\ra}[1]{\renewcommand{\arraystretch}{#1}}

\DeclareUnicodeCharacter{0301}{\'}

\title{Laugh Now Cry Later: Controlling Time-Varying Emotional States of Flow-Matching-Based Zero-Shot Text-to-Speech}

\name{\begin{tabular}{c}
    Haibin Wu$^{1\dagger}$, Xiaofei Wang$^2$, Sefik Emre Eskimez$^2$, Manthan Thakker$^2$, Daniel Tompkins$^2$,\\
 Chung-Hsien Tsai$^2$, Canrun Li$^2$, Zhen Xiao$^2$, Sheng Zhao$^2$, Jinyu Li$^2$, Naoyuki Kanda$^2$
\end{tabular}\thanks{$^\dagger$Work performed during an internship at Microsoft.}}
\address{$^1$National Taiwan University, $^2$Microsoft Corporation, USA}

\begin{document}
\ninept
\maketitle
\begin{abstract}

People change their tones of voice, often accompanied by nonverbal vocalizations (NVs) such as laughter and cries, to convey rich emotions. However, most text-to-speech (TTS) systems lack the capability to generate speech with rich emotions, including NVs. This paper introduces EmoCtrl-TTS, an emotion-controllable zero-shot TTS that can generate highly emotional speech with NVs for any speaker. EmoCtrl-TTS leverages arousal and valence values, as well as laughter embeddings, to condition the flow-matching-based zero-shot TTS. To achieve high-quality emotional speech generation, EmoCtrl-TTS is trained using more than 27,000 hours of expressive data curated based on pseudo-labeling.  Comprehensive evaluations demonstrate that EmoCtrl-TTS excels in mimicking the emotions of audio prompts in speech-to-speech translation scenarios. We also show that EmoCtrl-TTS can capture emotion changes, express strong emotions, and generate various NVs in zero-shot TTS. See \url{https://aka.ms/emoctrl-tts} for demo samples.

\end{abstract}

% \pagebreak

\input{sections/intro}

\input{sections/related}

\input{sections/method}

\input{sections/exp}

\input{sections/conclusion}

% \pagebreak
\bibliographystyle{IEEEbib}
\bibliography{strings,refs}

\end{document}

%% file: sections/intro.tex
\section{Introduction}
\label{sec:intro}
\vspace{-.5em}

Humans express a wide range of emotions by changing their tone of voice, often accompanied by nonverbal vocalizations (NVs) such as laughter and crying. While current emotional text-to-speech (TTS) systems have made significant advancements~\cite{zhao2023emotion,tang2024ed,guo2023emodiff,tang2023emomix,zhou2022speech,tang2023qi,lei2022msemotts,shin2022text,lee2017emotional,li2021controllable,cai2021emotion,cho2024emosphere},
% (comparison is shown in Table~\ref{tab:tts_comparison}), 
they still 
% can't 
lack the ability to 
generate emotional speech with fine-grained control (e.g. changing the emotion states within a single generated utterance) and with various types of NVs like laughter and crying. 
In addition, current emotional TTS systems~\cite{zhao2023emotion,tang2024ed,guo2023emodiff,tang2023emomix,zhou2022speech,tang2023qi,lei2022msemotts,shin2022text,lee2017emotional,li2021controllable,cai2021emotion,cho2024emosphere} are typically trained on staged datasets with a limited number of speakers; in extreme cases, some are trained on only one speaker. 
These TTS models often lack the ability to generate emotional speech for any speaker, a feature critical for applications like speech-to-speech translation, that needs to retain both the emotion and speaker characteristics of the source audio when generating the translated speech.

% Our contributions include the proposal of a framework that integrates flow-matching-based zero-shot TTS with NV and emotion embeddings, as well as comprehensive experiments to evaluate it, including the introduced new emotion-control evaluation metrics.
In this paper, we propose EmoCtrl-TTS, an emotion-controllable zero-shot TTS system that can generate highly emotional speech with NVs for any speaker. EmoCtrl-TTS generates the speech by mimicking the voice characteristics and emotion presented by an audio sample, referred to as an audio prompt.
EmoCtrl-TTS is based on the flow-matching-based zero-shot TTS~\cite{le2024voicebox} and utilizes valence and arousal values to mimic the time-varying characteristics of emotions.
In addition, it also utilizes laughter embeddings \cite{kanda2024making}, which we find to be effective for generating not only laughter but also other NVs, including crying. 
Furthermore, by leveraging over 27k hours of highly expressive real-world data through careful data mining, EmoCtrl-TTS achieves significant enhancements in robustness.
Comprehensive evaluations demonstrate that EmoCtrl-TTS excels in reproducing the emotions of audio prompts across multiple languages in speech-to-speech translation scenarios. 
We also show that EmoCtrl-TTS can capture emotion changes, express strong emotions, and generate various types of NVs in zero-shot TTS.
Our contributions are summarized as follows: (1) we propose a framework integrating flow-matching-based zero-shot TTS with NV and emotion embeddings; and (2) we conduct comprehensive experiments to evaluate the emotion-controllable zero-shot TTS, demonstrating the superiority of the proposed method.

%% file: sections/related.tex
\begin{figure*}[t]
    \centering
    \includegraphics[width=\textwidth]{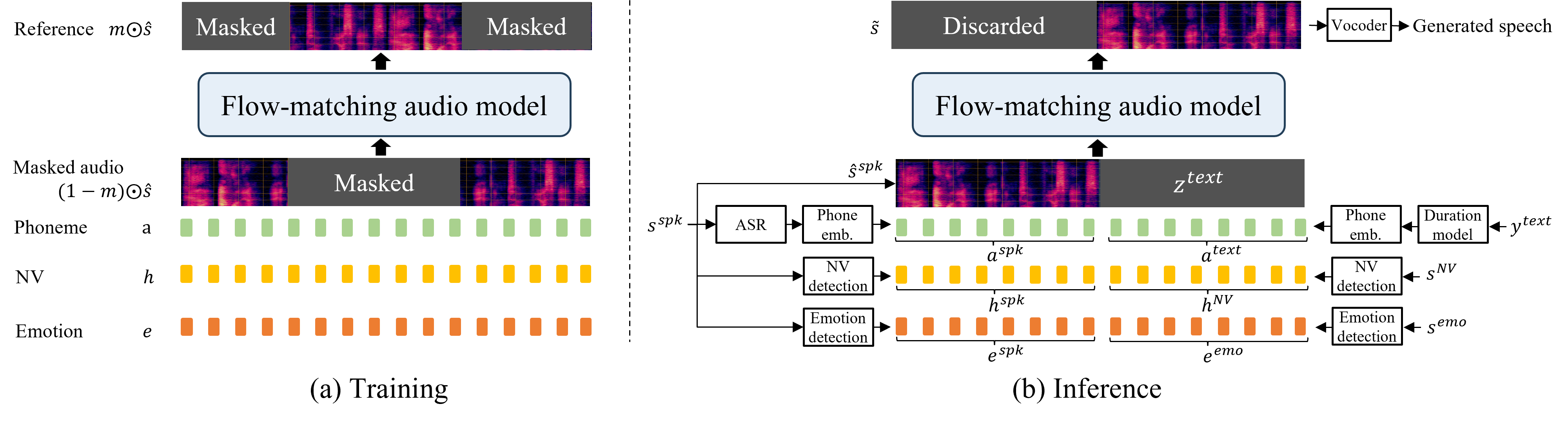}
    \vspace{-25pt}
    \caption{An overview of (a) training and (b) inference of the audio model of EmoCtrl-TTS. }
    \vspace{-15pt}
    \label{fig:overview}
\end{figure*}

\vspace{-.5em}
\section{Related Work}
\label{sec:relatedwork}

\vspace{-.5em}
\subsection{Controlling emotion in TTS}
\vspace{-.5em}

\input{tables/tts_comparison}

Emotional TTS has undergone substantial advancements in recent years. Table~\ref{tab:tts_comparison} lists various TTS systems from different perspectives.

The first point is whether TTS systems can control the fine-grained emotional attributes within one utterance. Such fine-grained control is ideal for many applications, for example, speech-to-speech translation where nuanced emotional changes need to be transferred to the translated speech. However, as in Table~\ref{tab:tts_comparison}, most prior works aimed to control the utterance-level emotion, and only a few works tackled the control of time-varying emotional status. MsEmoTTS~\cite{lei2022msemotts} used a local emotional strength predictor to estimate syllable-level emotion strength, leveraging it as a condition to control the emotion strength of generated speech. ELaTE~\cite{kanda2024making} leveraged laughter representation to condition the flow-matching-based zero-shot TTS, and showed superior controllability of laughter generation. However, they still lack full controllability of the emotional status.

The second point is the capability to generate NVs. As far as we investigated, most prior emotional TTS works were not able to generate NVs. While ELaTE~\cite{kanda2024making} can generate natural laughter, it was not investigated with other NVs such as cries. Our work aims to generate arbitrary types of NVs, including laughter and cries.

The third point is the size of the training data. Due to the difficulty in developing high-quality emotional training data with supervision, most works utilized less than 100 hours of training data. While ELaTE~\cite{kanda2024making} used 460 hours of speech containing laughter, the data scale is still less than 500 hours. To the best of our knowledge, ours is the first to investigate the impact of using large-scale emotional data for TTS training.

The fourth point is the number of speakers. As shown in the table, most of the emotional TTS systems utilized voices from fewer than 100 speakers, with some exceptions where the number of speakers is not available. While the number of speakers in our data is also unavailable due to anonymization, we expect that our training data contains a significantly large variation of speakers given the data scale, which is beneficial for the zero-shot TTS capability.

Finally, the fifth point is whether the emotional training data is staged data or real data. As shown in the table, most existing works utilized staged data for their training, which inevitably limited the variety of the speech. For example, in a speech-to-speech translation scenario, the source language speakers are often not professional actors, and their voice characteristics are different from the staged voice. By using large-scale training data, we aim to achieve highly faithful emotion transfer in the zero-shot TTS scenario.

\vspace{-.5em}
\subsection{Flow-matching-based TTS}
\vspace{-.5em}
\subsubsection{Conditional flow matching}
\vspace{-.5em}

Conditional flow-matching~\cite{lipman2023flow} is an objective for training generative models. 
Continuous normalizing flows~\cite{chen2018neural} is employed to convert a simple prior distribution $p_0$ into a complex distribution $p_1$ that aligns with the data.
To be specific, for a given data point $x$, a neural network parameterized by $\theta$ models a time-dependent vector field $v_t(x;\theta)$. 
This vector field constructs a flow $\phi_t$, which reshapes the prior distribution into the target distribution.
Lipman et al.~\cite{lipman2023flow} proposed to train such a neural network 
with the conditional flow-matching objective:
\begin{equation}
\mathcal{L}^{\rm CFM}(\theta)=\mathbb{E}_{t,q(x_1), p_t(x|x_1)}||u_t(x|x_1)-v_t(x;\theta)||^2,
\end{equation}
where $q$ denotes the training data distribution, $x_1$ represents the random variable for the training data, $p_t$ denotes the probability path at time step $t$, and $u_t$ is the vector field associated with $p_t$.
Also, Lipman et al.~\cite{lipman2023flow} present a conditional flow known as the optimal transport path, defined by the equations $p_t(x|x_1) = \mathcal{N}(x|tx_1, (1 - (1 - \sigma_{\rm min})t)^2I)$ and $u_t(x|x_1) = (x_1 - (1 - \sigma_{\rm min})x)/(1 - (1 - \sigma_{\rm min})t)$.

\vspace{-.5em}
\subsubsection{Voicebox}
\vspace{-.5em}

Voicebox \cite{le2024voicebox} is the first to utilize conditional flow matching for training zero-shot TTS. 
It is designed to perform speech-infilling tasks given audio context and frame-wise phoneme sequence as conditions.
Given the promising performance of Voicebox, we also employ conditional flow matching to develop our TTS model.

\vspace{-.5em}
\subsubsection{ELaTE}
\vspace{-.5em}
\label{sec:elate}

ELaTE~\cite{kanda2024making} was proposed to generate natural laughing speech with fine-grained controllability.
It utilized a frame-level laughter representation derived from 
the laughter detector~\cite{ryokai2018capturing, gillick2021robust}\footnote{\label{fn:laugh}\url{https://github.com/jrgillick/laughter-detection}} 
to condition the flow-matching-based zero-shot TTS, showing significantly higher quality and better controllability in generating laughing speech compared to conventional models.
However, ELaTE was only tested with laughing speech, and the effects on other NVs, such as crying, have not been investigated. 
In our preliminary experiment, we found that ELaTE sometimes generates laughing speech even when the audio prompt contains different NVs, such as crying. Our work can be regarded as an extension of ELaTE, where we aim to achieve better emotion controllability as well as the generation of various NVs.

%% file: tables/tts_comparison.tex
\begin{table}[t]
\caption{Comparison of TTS models based on emotion capabilities. 
}
\label{tab:tts_comparison}
\ra{0.9}
\centering
 \resizebox{\columnwidth}{!}{
    \footnotesize
     \tabcolsep = 0.4mm
    \begin{tabular}{@{}lccccc@{}}
    \toprule

    \multirow{2}{*}{\textbf{Model}} & \multirow{2}{*}{\textbf{\parbox[c]{1cm}{\centering Emotion\\change}}} & \multirow{2}{*}{\textbf{NVs}} & \multirow{2}{*}{\textbf{\parbox[c]{1.5cm}{\centering Emotion\\data size}}} & \multirow{2}{*}{\textbf{\parbox[c]{1cm}{\centering Speaker\\number}}} & \multirow{2}{*}{\textbf{\parbox[c]{2.0cm}{\centering Emotion\\data type}}} \\ 
    \\
    \midrule

    Emo-VITS~\cite{zhao2023emotion} & \ding{55} & \ding{55} & N/A & N/A & Staged \\ 
    ED-TTS~\cite{tang2024ed} & \ding{55} & \ding{55} & 70 hours & 1 & Staged \\ 
    EmoDiff~\cite{guo2023emodiff} & \ding{55} & \ding{55} & 12 hours & 10 & Staged \\
    EmoMix~\cite{tang2023emomix} & \ding{55} & \ding{55} & $\sim$15 hours & 10 & Staged \\
    Zhou et al.~\cite{zhou2022speech} & \ding{55} & \ding{55} & $\sim$30 hours & $\sim$100 & Staged \\ 
    QI-TTS~\cite{tang2023qi} & \ding{55} & \ding{55} & $\sim$15 hours & 10 & Staged \\
    MsEmoTTS~\cite{lei2022msemotts} & intensity & \ding{55} & 22 hours & 1 & Staged \\
    Shin et al.~\cite{shin2022text} & \ding{55} & \ding{55} & 57 hours & 38 & Staged \\
    Lee et al.~\cite{lee2017emotional} & \ding{55} & \ding{55} & 21 hours & 1 & Staged \\
    Li et al.~\cite{li2021controllable} & \ding{55} & \ding{55} & 14 hours & 1 & Staged \\
    Cai et al.~\cite{cai2021emotion} & \ding{55} & \ding{55} & 73 hours & 1 & Staged \\ % \midrule
    EmoSphere-TTS~\cite{cho2024emosphere} & \ding{55} & \ding{55} & $\sim$29 hours & 20 & Staged \\ 
    ELaTE~\cite{kanda2024making} & happy$\leftrightarrow$neutral & laugh & 460 hours & N/A & Real\\ \midrule

    EmoCtrl-TTS & arbitrary & arbitrary & $\sim$27k hours & N/A & Real \\ 
    \bottomrule
    \end{tabular}
}
% \vspace{-.5em}
\vspace{-5mm}
\end{table}

%% file: sections/method.tex
\vspace{-.5em}
\section{EmoCtrl-TTS}
\label{sec:method}

\vspace{-.5em}
\subsection{Overview}
\label{ssec: method overview}
\vspace{-.5em}

\subsubsection{Model training}
Figure~\ref{fig:overview} (a) illustrates the training procedure of EmoCtrl-TTS.
Given a training audio sample $s$ with transcription $y$, we extract its mel-filterbank features $\hat{s}\in\mathbb{R}^{F\times T}$, where $F$ denotes the feature dimension and $T$ represents the sequence length. 
Additionally, we employ force alignment and a phoneme embedding layer to obtain a frame-wise phoneme embedding $a \in\mathbb{R}^{D^{phn}\times T}$, where $D^{phn}$ is the phoneme embedding dimension. 
The phoneme embedding layer is a part of the audio model and is jointly trained.
Furthermore, we extract frame-wise embeddings that represent
NV  $h \in\mathbb{R}^{D^{NV}\times T}$ and emotion $e \in\mathbb{R}^{D^{emo}\times T}$,
where $D^{NV}$ and $D^{emo}$ denote the dimensions of NV and emotion embeddings respectively.
The embeddings $h$ and $e$ are extracted by using pre-trained NV and emotion detector,
respectively, which are discussed in Section \ref{ssec:nv_feature} and \ref{ssec:feature}.
We leverage the speech infilling task introduced in~\cite{le2024voicebox} to train the audio model, focusing on training a conditional flow-matching model to estimate the distribution $P(m\odot\hat{s}|(1-m)\odot\hat{s},a,h,e)$,
where $m\in\{0,1\}^{F\times T}$ represents a binary temporal mask, and $\odot$ 
is the Hadamard product.

\subsubsection{Inference}
Figure~\ref{fig:overview} (b) illustrates the inference procedure 
of EmoCtrl-TTS.
During inference, the model takes four inputs: 
text prompt $y^{text}$,
speaker prompt audio $s^{spk}$,
NV prompt audio $s^{NV}$, 
and emotion prompt audio $s^{emo}$.
The text prompt represents the content of the generated speech. Meanwhile, the speaker, 
NV, and emotion prompts control the characteristics of the speaker, NV, and emotion in the generated speech, respectively.
In speech-to-speech translation scenario, we use the source audio 
for $s^{spk}$, $s^{NV}$ and $s^{emo}$, and
translated text as $y^{text}$.
This results in the translated speech maintaining the source speaker’s voice and emotional characteristics.

The speaker prompt $s^{spk}$ is first converted to the mel-filterbank features $\hat{s}^{spk}$. It is also converted to phoneme embeddings $a^{spk}$ by applying automatic speech recognition (ASR) and then the phoneme embedding layer. The speaker prompt is further converted to NV embeddings $h^{spk}$ and emotion embeddings $e^{spk}$ based on the NV detector and emotion detector, respectively.

Meanwhile, the text prompt $y^{text}$ is converted to text prompt embeddings $a^{text}$ based on the phone duration model~\cite{le2024voicebox} followed by the phoneme embedding layer. The NV prompt embedding $h^{NV}$ and the emotion prompt embedding $e^{emo}$ are extracted from the NV detector and emotion detector, respectively. 
Note if the lengths of $h^{NV}$ and $h^{emo}$ are different from that of $a^{text}$, we apply linear interpolation to $h^{NV}$ and $h^{emo}$ to match their lengths to that of $a^{text}$.

The flow-matching-based audio model will then generate mel-filterbank features $\tilde{s}$ based on the learned distribution of $P(\tilde{s}|[\hat{s}^{spk}; z^{text}], [a^{spk}; a^{text}], [h^{spk}; h^{NV}], [e^{spk}; e^{emo}])$, where $z^{text}$ 
is an all-zero matrix with a shape of ${F\times T^{\rm text}}$,
and $[;]$ denotes concatenation operation in the time dimension. The generated part of $\tilde{s}$ is then converted to the speech signal using a vocoder.

\vspace{-.5em}
\subsection{NV embeddings}
\label{ssec:nv_feature}

For our proposed framework, it is essential to figure out a suitable embedding that can represent the characteristics of various NVs. In ELaTE~\cite{kanda2024making}, an embedding obtained from an off-the-shelf laughter detection model~\cite{ryokai2018capturing, gillick2021robust}\footref{fn:laugh} was used to control laughter in zero-shot TTS.

One of our findings in this work is that this laughter detector-based embedding actually captures a broader range of NV types than just laughter. By appropriately using the laughter-detector-based embedding, we have successfully generated various NVs such as crying and moaning.
Therefore, throughout this paper, we use a 32-dimensional embedding from the laughter detection model as the NV embedding.

\vspace{-.5em}
\subsection{Emotion embeddings}
\label{ssec:feature}
\vspace{-.5em}
Based on the Russell's circumplex model of emotion,
emotions can be represented in two major ways \cite{russell1980circumplex}:
Firstly, emotions can be categorized into different emotion classes, such as happiness or sadness, reflecting distinct emotional states. 
Secondly, emotions can be described using two attributes, arousal, and valence, sometimes with the third attribute of dominance. 
% Arousal refers to the intensity of the emotion, whether calm or highly stimulated. 
Arousal refers to the level of intensity or activation of the emotion, ranging from calm to highly stimulated.
% Valence indicates whether the emotion is positive or negative. 
Valence refers to how pleasant or unpleasant an emotion is, ranging from very positive to very negative.
Dominance relates to how much control one feels over the situation.

Finding an effective emotion embedding is crucial for our framework. 
In our preliminary experiment,
we used the eight emotion categories determined in \cite{lotfian2017building}
where each emotion category was represented by a learnable embedding,
which is then used as $e^{spk}$.
However, we found the TTS model struggled to generate emotionally expressive speech with this approach. % when provided with different emotion classes as inputs.
We also explored using the prosody encoder of the FACodec \cite{ju2024naturalspeech}. However, we found that the output of the prosody encoder contains phonetic information, resulting in the speech generation following the contents of the emotion prompt audio rather than the text prompt.
%the reduced intelligibility of the generated speech.

Ultimately, we identify a promising representation: arousal and valence values
predicted by a pre-trained arousal-valence-dominance extractor \cite{wagner2023dawn}\footnote{\label{fn:aro-val}\url{https://github.com/audeering/w2v2-how-to}}. 
This extractor is initialized with a wav2vec 2 model \cite{baevski2020wav2vec} and fine-tuned on MSP-PODCAST data \cite{lotfian2017building} to predict arousal, valence, and dominance values. 
Chunk-wise arousal-valence values ($D^{emo}=2$) are extracted using a sliding window with a window size of 0.5 seconds and a hop size of 0.25 seconds. 
Because the extractor outputs each value in the range of 0.0 to 1.0, we subtract 0.5 from the estimated value to adjust the range from -0.5 to 0.5.
We align the length of the extracted values with the phoneme embedding through linear interpolation.
This representation allows for capturing more nuanced emotional variations within each utterance.
Note that 
our preliminary investigations revealed that the additional use of the dominance value hurt the audio quality; therefore, we omitted the dominance value.

\vspace{-.5em}
\subsection{Collecting large-scale emotional data with pseudo-labeling}
\label{ssec:data}
\vspace{-.5em}

The quantity and quality of training data is a crucial factor for achieving high-quality TTS. 
However, either the recording of emotional speech or manual annotation of the recording is costly, making it difficult to scale the data size to more than 100 hours.

In this work, we curate 27k hours of highly emotional data, referred to as In-house Emotion Data (IH-EMO) in this paper, from 200k hours of in-house unlabeled anonymized English audio~\cite{wang2024investigation}. The data curation procedure is as follows. We first employ the emotion2vec model \cite{ma2023emotion2vec}\footnote{\label{fn:emo2vec}\url{https://github.com/ddlBoJack/emotion2vec}} to obtain predicted emotion confidence scores. We retain the samples if the predicted emotion is \{angry, disgusted, fearful, sad, surprised\} or the predicted emotion is \{neutral, happy\} with a confidence score of 1.0.\footnote{The dominant predictions were neutral or happy. To balance the emotion category, we decided to exclude the samples with low confidence scores for these two emotion classes. Even after this procedure, these two emotion classes were still the top-2 categories in the collected data.} We further apply DNSMOS~\cite{reddy2022dnsmos} and retain only samples whose OVLR score is greater than 3.0. Finally, we also apply an in-house speaker change detection model and discard the sample whenever a speaker change is detected. As a result, 27k hours of emotional audio are collected.
We use an off-the-shelf speech recognition model\footnote{\label{fn:kaldi}\url{https://kaldi-asr.org/models/m13}} to obtain the transcription.

%% file: sections/exp.tex
\vspace{-.5em}
\section{Experiments}
\vspace{-.5em}
\label{sec:experiments}

\input{sections/exp-1}

\input{tables/ablation_table}

\input{tables/real_data_table}

\input{sections/exp-4-result}

%% file: sections/exp-1.tex
\vspace{-.5em}
\subsection{Data}

\vspace{-.5em}
\subsubsection{Training data}

\vspace{-.5em}

We used three training datasets: Libri-light, LAUGH, and IH-EMO. Libri-light was used for pre-training the audio model without NV and emotion embeddings. On the other hand, LAUGH and IH-EMO were used for fine-tuning the audio model with NV and emotion embeddings. During fine-tuning, we also used the Libri-light data with a certain probability as suggested in \cite{kanda2024making}. The overview of each dataset is as follows.

\textbf{Libri-light~\cite{kahn2020libri}}: 60k hours of untranscribed English audiobooks from over 7,000 speakers. 
We transcribed the data by using a pre-trained Kaldi ASR model\footref{fn:kaldi}, 
which was trained on the 960-hour LibriSpeech dataset \cite{panayotov2015librispeech}.

\textbf{LAUGH}: 460 hours of laughing speech collected from the AMI meeting corpus \cite{carletta2005ami}, Switchboard corpus \cite{godfrey1992switchboard}, and Fisher corpus \cite{cieri2004fisher}. We gather all the utterances marked with laughter from the transcriptions of each corpus.  Note that the dataset still contains a substantial amount of neutral speech, as laughter tends to occur at certain parts of the speech.

\textbf{IH-EMO}: 27k hours of collected emotional speech as described in Section \ref{ssec:data}.

\vspace{-.5em}
\subsubsection{Evaluation data}
\vspace{-.5em}

We used four evaluation datasets as presented in Table~\ref{table:eval_datasets}.

\input{tables/evaluation_data}

\textbf{JVNV speech-to-speech translation (S2ST)}: To evaluate the emotion transferability of zero-shot TTS models, we established an experimental setting based on the Japanese-to-English S2ST scenario. In the evaluation, we used a Japanese speech from the JVNV corpus~\cite{xin2024jvnv}.
The JVNV corpus includes 
speeches from four speakers (two males and two females) with 
six emotions: anger, disgust, fear, happiness, sadness, and surprise with various NVs.
%, including not only laugher and crying, but also moaning. 
With its intense emotional expressions and various NVs, the JVNV dataset offers comprehensive coverage of expressive emotions, making it ideal for our emotion transfer testing. 
In the evaluation process,
we first applied speech-to-text translation to the Japanese speech to obtain the English translations. We then used a zero-shot TTS model, using the English translations as a text prompt and the Japanese speech as an audio prompt, from which we extracted both NV and emotion embeddings. The generated speech is expected to be English-translated speech with the original speaker's voice and emotional characteristics.
We assessed the similarity of the speaker and emotion between the source Japanese speech and the translated English speech using various metrics described in the next section.

\textbf{EMO-change}: To test the model's capacity for fine-grained emotional speech generation, we created the EMO-change dataset based on the RAVDESS dataset \cite{livingstone2018ryerson}, which contains English emotional speech data with emotions such as calm, happy, sad, angry, fearful, surprised, and disgusted. The RAVDESS dataset includes two transcriptions: ``kids are talking by the door" and ``dogs are sitting by the door" with intense emotional expressions. 
To generate the EMO-change dataset, we randomly select two utterances with different emotions with the transcription ``kids are talking by the door," remove the silence from each, and concatenate them to create the audio prompts. During testing, the concatenated emotion change sample serves as the audio prompt, and a repeated sentence ``dogs are sitting by the door dogs are sitting by the door" serves as the text prompt for the zero-shot TTS model. This setup evaluates the model's ability to generate speech that mimics the emotional transitions in the audio prompt while following the given text prompt, and speaker characteristics.

\textbf{Laughter-test}: To evaluate the zero-shot TTS capability in generating laughing speech, we employed a Chinese-to-English S2ST experiment
by following the protocol presented in~\cite{kanda2024making}.
Specifically, 
we used 154 Chinese utterances containing laughter\footnote{Data can be found at \url{https://aka.ms/elate}.}
from the evaluation subset of the DiariST-AliMeeting test set \cite{yang2024diarist}.
We applied zero-shot TTS, where
we used the ground-truth English transcription as the text prompt and 
the Chinese speech as the audio prompt to extract both 
NV and emotion embeddings. 
The generated speech is expected to be English speech with the original speaker's voice and laughter characteristics.

\textbf{Crying-test}: 
To
evaluate the zero-shot TTS capability in generating crying speech, 
we further conducted a Chinese-to-English S2ST experiment
using 33 Chinese crying speech samples collected from publicly available data sources.
We followed the same procedure with the Laughter-test,
and the generated speech is expected to be English speech with the original speaker's voice and crying characteristics.

\vspace{-.5em}
\subsection{Evaluation metrics}
\vspace{-.5em}

\subsubsection{Objective evaluation metrics}
\vspace{-.5em}

We utilized the following objective metrics. Among them, AutoPCP, EMO SIM, and Aro-Val SIM are closely related to the emotion controllability.\footnote{\textcolor{black}{\url{https://github.com/hbwu-ntu/EmoCtrlTTS-Eval}}}

\textbf{Word error rate (WER):} To evaluate the intelligibility of the generated audio, 
we applied a Whisper-Large~\cite{radford2023robust} to the generated audio and computed the WER. 
In all our tables, we express the WER in percentage.

\textbf{Speaker SIM-o:} To evaluate the speaker similarity between the generated audio and the audio prompt, we computed a cosine similarity between speaker embeddings of the two audios. We used a WavLM-large-based 
speaker verification model~\cite{chen2022wavlm}\footnote{\url{https://github.com/microsoft/UniSpeech/tree/main/downstreams/speaker\_verification}}, following previous works \cite{wang2023neural,le2024voicebox}.

\textbf{AutoPCP:} AutoPCP \cite{barrault2023seamless} is an utterance-level estimator to quantify the prosody similarity between two speech samples.
We computed the score between the generated audio and the audio prompt.
We leveraged AutoPCP\_multilingual\_v2\footnote{\url{https://github.com/facebookresearch/seamless\_communication}}.

\textbf{Emo SIM}: To evaluate the similarity of time-varying emotion states, we applied the emotion2vec model~\cite{ma2023emotion2vec}\footref{fn:emo2vec} to extract the emotion embeddings. 
We performed interpolation to ensure the embeddings of the audio prompt and the generated audio have the same length. We then computed the cosine similarity between these two embedding sequences for each frame and took the average to obtain the EMO SIM score.
 
\textbf{Aro-Val SIM}: As another metric for the similarity of time-varying emotion states, we computed the arousal-valence values based on \cite{wagner2023dawn}\footref{fn:aro-val} using a sliding window with a window size of 0.5 sec and a hop size of 0.25 sec. Similar to EMO SIM, we computed the cosine similarity between the audio prompt and the generated audio for every frame and took the average as the Aro-Val SIM.

\vspace{-.5em}
\subsubsection{Subjective evaluation metrics}
\vspace{-.5em}
We used the following subjective evaluation metrics.

\textbf{SMOS}: Speaker similarity mean opinion score, which is the similarity between the speaker prompt and the generated speech from 1 (not at all similar) to 5 (extremely similar).
    
\textbf{NMOS}: Naturalness MOS, which is the naturalness of the generated speech from 1 (bad) to 5 (excellent).
    
\textbf{EMOS}: Emotion MOS, which measures the similarity of emotion between the audio prompt and the generated speech from 1 (not at all similar) to 5 (extremely similar).

\input{tables/jvnv_ravdess_table}

\input{tables/subj_eval_table}

\vspace{-.5em}
\subsection{Model configuration}
\vspace{-.5em}

The architecture of the EmoCtrl-TTS audio model closely followed the configurations of the Voicebox \cite{le2024voicebox}. Specifically, we used a Transformer~\cite{vaswani2017attention} with 24 layers, featuring 16 attention heads, a 1024-dimensional embedding, and a feed-forward layer with a dimension of 4096.

The model was pre-trained using Libri-light data, without NV or emotion embedding. This resulted in the reproduction of the Voicebox model (B2 in Table~\ref{tab:jvnv_results}). The model was trained for 390K steps with an effective mini-batch size of 307,200 audio frames.
A linear-decay learning rate scheduler with a peak learning rate at 7.5e-5 was used, along with 20K steps of linear warmup.
After the pre-training, we further fine-tuned the model by combining Libri-light, LAUGH, and IH-EMO.
During fine-tuning, 
the effective mini-batch size was set to 307,200 audio frames. 
A linear-decay learning rate scheduler was used with a peak learning rate of 7.5e-5.
Unless otherwise stated, we fine-tuned the model with 40k steps.

During the inference, we used classifier-free guidance with a guidance strength of 1.0, and the number of function evaluations was set to 32. 
A MelGAN-based vocoder~\cite{kumar2019melgan} was used to convert the mel spectrogram into waveforms.

\vspace{-.5em}
\subsection{S2ST pipeline} % for JVNV S2ST, Laughter-test and Crying-test}
\vspace{-.5em}

For the JVNV S2ST evaluation data, we leveraged the Whisper large-v3 model~\cite{radford2023robust} to transcribe the Japanese utterances with time stamps. We then employed GPT-4~\cite{achiam2023gpt} to translate the time-stamped Japanese text into English by keeping the time stamps.
We then used a total-duration-aware (TDA) duration model~\cite{eskimez2024total} to obtain a frame-wise phoneme alignment.
For the Laughter-test and Crying-test, we used the ground-truth English text translation,
and the same TDA duration model
to obtain the frame-wise phoneme alignment.

%% file: tables/evaluation_data.tex
\begin{table}[t]
\centering
\caption{Summary of evaluation datasets. S2ST: Speech-to-speech translation.}
\label{table:eval_datasets}
\resizebox{0.48\textwidth}{!}
{
%\tiny
\centering
 \tabcolsep = 0.5mm
    \begin{tabular}{@{}ccccc@{}}
    \toprule
    \textbf{Dataset} & \textbf{Source Language} & \textbf{Utterances} & \textbf{Type} & \textbf{Task} \\ 
    \midrule
%    \hdashline[1pt/2pt]\hdashline[0pt/1pt]
    JVNV S2ST & Japanese (ja) & 1615 & Staged & S2ST (ja$\rightarrow$en) with various emotions \\ 
    %\hdashline[1pt/2pt]\hdashline[0pt/1pt]
    EMO-change & English (en) & 84 & Simulated & Zero-shot TTS with emotion change \\ %within the utterance \\ 
    %\hdashline[1pt/2pt]\hdashline[0pt/1pt]
    Laughter-test & Chinese (zh) & 154 & Real & S2ST (zh$\rightarrow$en) with laughter\\ 
    %\hdashline[1pt/2pt]\hdashline[0pt/1pt]
    Crying-test & Chinese (zh) & 33 & Real & S2ST (zh$\rightarrow$en) with crying \\ \bottomrule
    \end{tabular}
}
\vspace{-3mm}
\end{table}

%% file: tables/jvnv_ravdess_table.tex
\begin{table*}[!t]
\caption{Objective evaluation results for various models on JVNV S2ST and EMO-change test sets.
A model with $^{(+)}$ was fine-tuned with 200k steps with more exposure to the IH-EMO. 
LL: Libri-light.}
\label{tab:jvnv_results}
\ra{0.9}
\centering
%\resizebox{\textwidth}{!}
{
\footnotesize
\tabcolsep = 0.7mm
%\centering
\begin{tabular}{@{}llccccccccc@{}}
    \toprule
    \multirow{1}{*}{\textbf{ID}} &
    \multirow{1}{*}{\textbf{Model}}  &
    \multirow{1}{*}{\textbf{Init.}} &
    $h$ &
    $e$ &
    \multirow{1}{*}{\textbf{Training Data (hours)}}  &
    \textbf{SIM-o$\uparrow$} & 
    \textbf{WER (\%)$\downarrow$} & 
    \textbf{AutoPCP$\uparrow$} & 
    \textbf{Emo SIM$\uparrow$} & 
    \textbf{Aro-Val SIM$\uparrow$} \\
    
    \midrule
    \multicolumn{11}{c}{\textbf{JVNV S2ST}} \\
    \midrule
    \textbf{(B1)} & SeamlessExpressive~\cite{barrault2023seamless} & - &- &- & -   & 0.268  & {\bf 1.2}  & 2.91 & 0.653 & 0.494  \\
    \textbf{(B2)} & Voicebox (reproduction)~\cite{le2024voicebox}  & - &- &-& LL (60k) & 0.347  & 2.1  & 2.96 & 0.655 & 0.443 \\
    \textbf{(B3)} & ELaTE~\cite{kanda2024making} & B2 &\checkmark &-& LL (60k) + LAUGH (460) & 0.441  & 3.8  & 3.36 & 0.671 & 0.548  \\
    \hdashline[1pt/2pt]\hdashline[0pt/1pt]
    \textbf{(B4)} & Voicebox (fine-tuned) & B2 &- &-& LL (60k) + LAUGH (460) & 0.410  & 2.5  & 3.07 & 0.645 & 0.438  \\
    \textbf{(B5)} & Voicebox (fine-tuned)       & B2&- &- & LL (60k) + IH-EMO (27k) & 0.479 & 2.2 & 2.96 & 0.641 & 0.397   \\
    \textbf{(B6)} & Voicebox (fine-tuned)      & B2&- &- & LL (60k) + IH-EMO (27k) + LAUGH (460) & 0.455 & 3.0 & 3.17 & 0.659 & 0.470  \\
    \hdashline[1pt/2pt]\hdashline[0pt/1pt]
    \textbf{(P1)} & EmoCtrl-TTS              & B2&\checkmark &\checkmark & LL (60k) + IH-EMO (27k) + LAUGH (460) & 0.448 & 4.4 & 3.38 & 0.693 & {\bf 0.647} \\
    \textbf{(P2)} & EmoCtrl-TTS$^{(+)}$ & B2&\checkmark &\checkmark & LL (60k) + IH-EMO (27k) + LAUGH (460) & {\bf 0.497} & 3.2 & {\bf 3.50} & {\bf 0.697} & 0.643 \\
    \midrule
    \multicolumn{11}{c}{\textbf{EMO-change}} \\
    \midrule
    \textbf{(B2)} & VoiceBox (reproduction)~\cite{le2024voicebox} & - &- &- & LL (60k) & 0.600 & 1.2 & 3.31 & 0.685 & 0.663 \\
    \textbf{(B3)} & ELaTE~\cite{kanda2024making}  & B2 &\checkmark &- & LL (60k) + LAUGH (460)    & 0.643 & 0.2 & {\bf 3.52} & {\bf 0.700} & 0.761 \\

    \hdashline[1pt/2pt]\hdashline[0pt/1pt]
    \textbf{(B6)} & Voicebox (fine-tuned)        & B2 &- &-& LL (60k) + IH-EMO (27k) + LAUGH (460) & 0.622 & 1.1 & 3.31 & 0.678 & 0.655  \\
    \textcolor{black}{\textbf{(P1)}} & \textcolor{black}{EmoCtrl-TTS}                  & B2 & \checkmark &\checkmark & LL (60k) + IH-EMO (27k) + LAUGH (460) & 0.671 & {\bf 0.0}  & 3.45 & 0.685 & {\bf 0.822}  \\
    \textbf{(P2)} & EmoCtrl-TTS$^{(+)}$ & B2 &\checkmark &\checkmark& LL (60k) + IH-EMO (27k) + LAUGH (460) & {\bf 0.684} & 0.9 & 3.44 & 0.679 & 0.811  \\

    \bottomrule
\end{tabular}
}
\vspace{-5mm}
\end{table*}

%% file: tables/subj_eval_table.tex
\begin{table}[t]
        \caption{Subjective evaluation results on the JVNV S2ST test set are presented. The group with the top score (scores within the 95\% confidence interval of the highest score) is displayed in \textbf{bold} font.}
    \label{tab:subj_results_jvnv}
    \ra{0.9}
   % \resizebox{\columnwidth}{!}
    {
    \footnotesize
        \begin{tabular}{@{}lllll@{}}
            \toprule
            \textbf{ID}    & \textbf{Model}  & \textbf{SMOS} & \textbf{NMOS} & \textbf{EMOS} \\ 
            \midrule 
            \multicolumn{5}{c}{\bf JVNV S2ST}\\
            \midrule
            \textbf{(B1)} & SeamlessExpressive~\cite{barrault2023seamless}         &  2.40$_{\pm 0.18}$  & 2.83$_{\pm 0.16}$  & 3.46$_{\pm 0.17}$    \\
            \textbf{(B2)} & Voicebox (repro.)~\cite{le2024voicebox}$^\dagger$  &  3.12$_{\pm 0.21}$  & 3.28$_{\pm 0.15}$  & 3.51$_{\pm 0.16}$    \\
            \textbf{(B3)} & ELaTE~\cite{kanda2024making}                       &  \bf{3.62}$_{\pm 0.20}$  & \textbf{3.43$_{\pm 0.14}$}  & \textbf{3.68$_{\pm 0.15}$}    \\
            \textbf{(B6)} & Voicebox (fine-tuned)                              &  \textbf{3.76$_{\pm 0.18}$}  & 3.29$_{\pm 0.14}$  & \textbf{3.67$_{\pm 0.16}$}    \\
            \textbf{(P2)} & EmoCtrl-TTS                                        &  \textbf{3.72$_{\pm 0.18}$}  & \textbf{3.53$_{\pm 0.13}$}  & \textbf{3.66$_{\pm 0.15}$}    \\ 
            \midrule
            \multicolumn{5}{c}{\bf EMO-change}\\
            \midrule
            \textbf{(B3)} & ELaTE~\cite{kanda2024making} &  4.37$_{\pm 0.09}$    &  \textbf{3.56}$_{\pm 0.12}$  &  \textbf{3.65}$_{\pm 0.13}$    \\
            \textbf{(B6)} & Voicebox (fine-tuned)            &  4.36$_{\pm 0.10}$    &  \textbf{3.53}$_{\pm 0.13}$    &  2.95$_{\pm 0.13}$    \\
            \textbf{(P2)} & EmoCtrl-TTS                   &  \textbf{4.49}$_{\pm 0.08}$    &  \textbf{3.55}$_{\pm 0.12}$  &  \textbf{3.54}$_{\pm 0.14}$    \\ 
            \bottomrule
        \end{tabular}
}
\vspace{-3mm}
\end{table}

%% file: tables/ablation_table.tex
\begin{table*}[t]
  \centering
      \caption{Impact of training configurations on the EmoCtrl-TTS on JVNV S2ST test set.} 
  \label{tab:ablation_study}
\ra{0.9}
\resizebox{\textwidth}{!}
{
% \scriptsize
\footnotesize
 \tabcolsep = 0.8mm
\begin{tabular}{@{}lccccccccccccccccc@{}}
    \toprule
    \multirow{2}{*}{\textbf{Training Data (hours)}} & \multirow{2}{*}{\textbf{Data Ratio}} & \multirow{2}{*}{\textbf{Steps}} && \multicolumn{2}{c}{\textbf{IH-EMO}} & \multicolumn{2}{c}{\textbf{LAUGH}} && \multirow{2}{*}{\textbf{SIM-o$\uparrow$}} && \multirow{2}{*}{\textbf{WER (\%)$\downarrow$}} && \multirow{2}{*}{\textbf{AutoPCP$\uparrow$}} && \multirow{2}{*}{\textbf{Emo SIM$\uparrow$}} && \multirow{2}{*}{\textbf{Aro-Val SIM$\uparrow$}}  \\
    % \cmidrule{7-15}
    & && & $h$ & $e$ & $h$ & $e$ &&  \\
    \midrule
     LL (60k) + LAUGH (460) & 0.5:0.5 & 40k           && && &&& 0.410  && 2.5  && 3.07 && 0.645 && 0.438  \\ 
     LL (60k) + LAUGH (460) (= ELaTE) & 0.5:0.5 & 40k && && \checkmark & && 0.441  && 3.8  && 3.36 && 0.671 && 0.548  \\
     \textcolor{black}{LL (60k) + LAUGH (460)} & 0.5:0.5 & 40k && && & \checkmark && 0.332 && 4.7 && 3.22 &&  0.680 &&  0.632 \\
     LL (60k) + LAUGH (460) & 0.5:0.5 & 40k           && && \checkmark & \checkmark && 0.391  && 4.4  && 3.39 && 0.702 && {\bf 0.663}  \\
    \hdashline[1pt/2pt]\hdashline[0pt/1pt]
     LL (60k) + IH-EMO (27k) & 0.5:0.5 & 40k && && &&& 0.479 && {\bf 2.2} && 2.96 && 0.641 && 0.397   \\
     \textcolor{black}{LL (60k) + IH-EMO (27k)} & 0.5:0.5 & 40k && \checkmark &  & &&& 0.403 && 6.7 && 3.14 && 0.687 &&  0.590  \\
     LL (60k) + IH-EMO (27k) & 0.5:0.5 & 40k && & \checkmark & &&& 0.487 && 2.5 && 3.36 && 0.682 && 0.608   \\
     LL (60k) + IH-EMO (27k) & 0.5:0.5 & 40k && \checkmark & \checkmark & &&& 0.438 && 7.7 && 3.29 && {\bf 0.707} && 0.637  \\
    \hdashline[1pt/2pt]\hdashline[0pt/1pt]
     LL (60k) + IH-EMO (27k) + LAUGH (460) & 0.5:0.25:0.25 & 40k & &&& &&& 0.455 && 3.0 && 3.17 && 0.659 && 0.470  \\
     LL (60k) + IH-EMO (27k) + LAUGH (460) & 0.5:0.25:0.25 & 40k && & \checkmark & \checkmark &&& 0.448 && 4.4 && 3.38 && 0.693 && 0.647  \\
    % \hdashline[1pt/2pt]\hdashline[0pt/1pt]
     LL (60k) + IH-EMO (27k) + LAUGH (460) & 0.5:0.4:0.1   & 40k && & \checkmark & \checkmark &&& 0.487 && 3.4 && 3.44 && 0.690 && 0.626  \\
     LL (60k) + IH-EMO (27k) + LAUGH (460) & 0.5:0.4:0.1   & 200k && & \checkmark & \checkmark &&& {\bf 0.497} && 3.2 && {\bf 3.50} && 0.697 && 0.643  \\
    \bottomrule
\end{tabular}
}
\vspace{-5mm}
\end{table*}

%% file: tables/real_data_table.tex
\begin{table}[t]
\vspace{-3mm}
\caption{Results of Laughter-test dataset. }
\label{tab:laughter_test_results}
\centering
\resizebox{\columnwidth}{!}
{
\scriptsize
\tabcolsep = 0.5mm
\begin{tabular}{@{}llccccc@{}}
    \toprule
    \multirow{1}{*}{\textbf{ID}} & \multirow{1}{*}{\textbf{Model}} & \textbf{SIM-o$\uparrow$} & \textbf{WER$\downarrow$} & \textbf{AutoPCP$\uparrow$} & \textbf{Emo SIM$\uparrow$} & \textbf{Aro-Val SIM$\uparrow$} \\
    \midrule
    \textbf{(B1)} & SeamlessExpressive~\cite{barrault2023seamless}   & 0.210  & 11.4 & 2.31 & 0.587 & 0.248  \\
    \textbf{(B2)} & Voicebox (repro.)~\cite{le2024voicebox}$^\dagger$  & 0.328  & 7.8  & 2.46 & 0.634 & 0.410   \\
    \textbf{(B3)} & ELaTE~\cite{kanda2024making}              & {\bf 0.399}  & 7.8  & {\bf 3.44} & 0.806 & 0.700   \\
    \textbf{(B6)} & Voicebox (fine-tuned)                         & 0.383  & {\bf 6.6}  & 2.47 & 0.689 & 0.410   \\
    \textbf{(P2)} & EmoCtrl-TTS                               & 0.392  & 9.9  & 3.38 & {\bf 0.848} & {\bf 0.795}   \\   
    \bottomrule
\end{tabular}
}
\vspace{-5mm}
\end{table}

\begin{table}[t]
\caption{Results of Crying-test dataset. }
\label{tab:crying_test_results}
\centering
\resizebox{\columnwidth}{!}
{
\scriptsize
\tabcolsep = 0.5mm
\begin{tabular}{@{}llccccc@{}}
    \toprule
    \multirow{1}{*}{\textbf{ID}} & \multirow{1}{*}{\textbf{Model}} & \textbf{SIM-o$\uparrow$} & \textbf{WER$\downarrow$} & \textbf{AutoPCP$\uparrow$} & \textbf{Emo SIM$\uparrow$} & \textbf{Aro-Val SIM$\uparrow$} \\
    \midrule
    \textbf{(B1)} & SeamlessExpressive~\cite{barrault2023seamless}   & 0.258  & 8.9  & 2.77 & 0.576 & 0.378  \\
    \textbf{(B2)} & Voicebox (repro.)~\cite{le2024voicebox}$^\dagger$  & 0.294  & 5.5  & 2.71 & 0.569 & 0.367  \\
    \textbf{(B3)} & ELaTE~\cite{kanda2024making}              & 0.384  & 5.6  & 3.14 & 0.642 & 0.471  \\
    \textbf{(B6)} & Voicebox (fine-tuned)                         & 0.383  & {\bf 5.1}  & 2.81 & 0.589 & 0.413  \\
    \textbf{(P2)} & EmoCtrl-TTS                               & {\bf 0.408}  & 7.5  & {\bf 3.21} & {\bf 0.662} & {\bf 0.597}  \\   
    \bottomrule
\end{tabular}
}
\vspace{-5mm}
\end{table}

%% file: sections/exp-4-result.tex
\vspace{-.5em}
\subsection{Results and discussion}
\label{sec:results}
\vspace{-.5em}

\subsubsection{Objective evaluation}
\vspace{-.5em}

Table ~\ref{tab:jvnv_results} presents the objective evaluation results on the JVNV and EMO-change datasets. During the evaluation, we generated speech
with three different random seeds, and the average of the scores was taken to report.

{\bf Baseline analysis:} 
For the JVNV S2ST test set, we first observed that the ELaTE (B3), trained with LL and LAUGH data, achieved superior performance over the SeamlessExpressive model~\cite{barrault2023seamless}\footnote{Our experiment is based on SeamlessExpressive, supported by the Seamless Licensing Agreement. Copyright © Meta Platforms, Inc. All Rights Reserved.} (B1) and the reproduced VoiceBox model (B2) across all the metrics, except for WER.
The comparison between B2 and B4  highlights the effect of LAUGH data, showing that the latter improved performance in SIM-o and AutoPCP while maintaining WER, EMO SIM, and Aro-Val SIM.
By replacing LAUGH with the IH-EMO data (B4 vs. B5), SIM-o and WER were improved from 0.410 to 0.479, and from 2.5\% to 2.2\%, respectively. However, this came at the cost of a slight degradation of AutoPCP (3.07$\rightarrow$2.96),  Emo SIM (0.645$\rightarrow$0.641), and Aro-Val SIM (0.438$\rightarrow$0.397). 
Combining all training data (B6) gave us a balanced result.
However, even after these trials, no Voicebox model achieved better AutoPCP, EmoSIM, 
and Aro-Val SIM compared to ELaTE.
The same trends were also observed for the EMO-change test set.
These results suggest that simply adding emotional data does not improve the
emotion transferability of the zero-shot TTS models.

{\bf Results of EmoCtrl-TTS:}
From JVNV S2ST results, we first observed (P1) showed a significant improvement in AutoPCP, Emo SIM, and Aro-Val SIM compared to the baselines.
By executing a longer fine-tuning (P2), EmoCtrl-TTS 
achieved further better SIM-o, WER, AutoPCP, and Emo SIM while almost preserving the high Aro-Val SIM.
For the EMO-change test set, 
EmoCtro-TTS achieved the best SIM-o and Aro-Val SIM 
and the second-best AutoPCP.
It is also noteworthy that SIM-o score was significantly improved 
from (B6) to (P1), which suggests the effectiveness of the use of NV and emotion embeddings not only for the emotion-related metrics but also for the speaker similarity.
Note that we have trained more variants of EmoCtrl-TTS to analyze the impact of
each data and each embedding, which will be discussed in Section \ref{ssec:train-config} with Table \ref{tab:ablation_study}.

\vspace{-.5em}
\subsubsection{Subjective evaluation}
\vspace{-.5em}

We performed the subjective evaluation for selected models.
We randomly picked 24 samples from JVNV S2ST data and 28 samples from EMO-change data, and asked native English testers to rate SMOS, NMOS and EMOS.
For JVNV S2ST test set, each sample was judged by 9, 11, and 10 testers for SMOS, NMOS, and EMOS, respectively. For EMO-change test set, each sample was judged by 12 testers for all metrics. 

The results are presented in Table ~\ref{tab:subj_results_jvnv}. 
From the JVNV S2ST result, the Voicebox model (B6), leveraging the IH-EMO data, demonstrated significant improvements of SMOS and EMOS from the vanilla Voicebox model (B2), which illustrates the importance of the training data. Meanwhile, ELaTE (B3) excelled in NMOS, suggesting that the integration of NV features significantly boosts the naturalness of the audio by generating an appropriate NV.
Finally, EmoCtrl-TTS (P2) achieved even better NMOS while keeping the high SMOS and EMOS scores.
%the best NMOS and comparable SMOS and EMOS to other top baselines, showcasing the superiority of our proposed approach.

In the EMO-change dataset, 
Voicebox (B6) showed significantly worse EMOS
compared to ELaTE (B3) and EmoCtlr-TTS (P2).
This result demonstrated that 
the conventional zero-shot TTS 
cannot mimic the time-varying
emotional states presented in the 
audio prompt.
The proposed EmoCtrl-TTS (P2) 
achieved best SMOS and
comparable NMOS and EMOS to other top 
baselines, again showcasing the
efficacy of the proposed approach.

\vspace{-.5em}
\subsubsection{Impact of training configurations}
\label{ssec:train-config}
\vspace{-.5em}

Table \ref{tab:ablation_study} presents the results of the JVNV S2ST test set with various training data configurations. 

The first four rows compare the performance of models trained by Libri-light and LAUGH data with a 0.5:0.5 ratio. By comparing the first two rows, we observe that the NV embedding $h$  significantly improved SIM-o, AutoPCP, Emo SIM, and Aro-Val SIM with the expense of degradation of WER. By introducing emotion embedding $e$, AutoPCP, EmoSIM, and Aro-Val SIM were further improved. However, it came with a cost of further degradation of WER as well as a significant drop in SIM-o.
    
In the 5th to 9th rows, we trained models using the Libri-light and IH-EMO data. Similar to the case with LAUGH data, we observed that the emotion embedding $e$ improved all the emotion-related metrics while keeping SIM-o and WER nearly intact. However, in contrast to the results with the LAUGH data, we observed a severe degradation in WER when the NV embedding $h$ was introduced. Upon examination, we found that the NV embedding often resulted in unwanted NV generation, such as laughter from multiple speakers.

Both observations suggest that including both NV and emotion embeddings in model training is not always beneficial. Based on these findings, we opted to use only emotion embeddings for IH-EMO and only NV embeddings for LAUGH when combining all training data.

The last four rows in Table \ref{tab:ablation_study} show the impact of data ratio and training steps on model performance with the combination of all training data.
We first observed that
the data ratio of 0.5:0.4:0.1 for Libri-light, IH-EMO, and LAUGH provided
us with a better SIM-o and WER than the ratio of 0.5:0.25:0.25
while keeping the emotion-related metrics similar. 
By fine-tuning the model longer,
further marginal improvement was obtained for all evaluation metrics.

\vspace{-.5em}
\subsubsection{Results on real laughter and crying data}
\vspace{-.5em}

Table ~\ref{tab:laughter_test_results} and ~\ref{tab:crying_test_results} present a comparison between the EmoCtrl-TTS model and selected baselines on the Laughter-test and Crying-test datasets, both of which are composed of real data. Compared to the baselines, the EmoCtrl-TTS consistently achieved
the best performance for Emo SIM and Aro-Val SIM while achieving either the best or the second-best Auto PCP and SIM-o.
It demonstrated EmoCtrl-TTS's robustness for the real data. 
On the other hand, we observe a moderate degradation of WERs. 
% We speculate the degradation would be attributed to the ASR model's limited ability to accurately recognize highly emotional speech.

%% file: sections/conclusion.tex
\section{Conclusions}
\label{sec:conclusions}

In this work, we presented EmoCtrl-TTS, an emotion-controllable zero-shot TTS model 
that can generate highly emotional speech with NVs for any speaker.
EmoCtrl-TTS leveraged arousal and valence values as well as the laughter embeddings to control the time-varying characteristics of emotional speech including NVs. 
Our comprehensive experiments demonstrated that EmoCtrl-TTS can closely mimic the voice characteristics and nuances of the source audio prompts by generating emotional speech with NVs.